\documentclass[a4paper,11pt]{article}
\usepackage[dvips]{graphicx}
\newcommand{\be}{\begin{equation}}
\newcommand{\en}{\end{equation}}
\newcommand{\bea}{\begin{eqnarray}}
\newcommand{\ena}{\end{eqnarray}}
\newcommand{\lbl}[1]{\label{eq:#1}}
\newcommand{\rf}[1]{(\ref{eq:#1})}
\newcommand{\np}[1]{Nucl.\ Phys.\ {\bf #1}}

\newcommand{\pr}[1]{Phys.\ Rev.\ {\bf #1}}
\newcommand{\prl}[1]{Phys.\ Rev.\ Lett.\ {\bf #1}}
\newcommand{\pl}[1]{Phys.\ Lett.\ {\bf #1}}
\newcommand{\ap}[1]{Ann. Phys. (NY)\ {\bf #1}}

\newcommand{\zp}[1]{Zeit.\ Phys.\ {\bf #1}}
\def\today{\ifcase \month\or
  January\or February\or March\or April\or May\or June\or
    July\or August\or September\or October\or November\or December\fi
      \space\number\day,\space
        \number\year }
\newcommand{\lapprox}{%
\mathrel{%
\setbox0=\hbox{$<$}\raise0.6ex\copy0\kern-\wd0\lower0.65ex\hbox{$\sim$}}}

\newcommand{\gapprox}{%
\mathrel{%
\setbox0=\hbox{$>$}\raise0.6ex\copy0\kern-\wd0\lower0.65ex\hbox{$\sim$}}}

\setbox0=\hbox{$\scriptstyle\circ$}
\newcommand{\Mchir}{
\mathrel{
\setbox1=\hbox{$M$}
\copy1\kern-0.5\wd1\raise1.1\ht1\copy0 }}

\newcommand{\Gchir}{
\mathrel{
\setbox1=\hbox{$\Gamma$}
\copy1\kern-0.7\wd1\raise1.1\ht1\copy0}}

\newcommand{\Pichir}{
\mathrel{\setbox0=\hbox{$\scriptstyle\circ$}
\setbox1=\hbox{$\Pi$}
\copy1\kern-0.7\wd1\raise1.1\ht1\copy0\kern0.125\wd1}}

\newcommand{\rochir}{
\mathrel{
\setbox1=\hbox{$\rho$}
\copy1\kern-0.6\wd1\raise1.1\ht1\copy0}}

\newcommand{\Dslash}{
\mathrel{
\setbox1=\hbox{$D$}
\setbox2=\hbox{$/$}
\copy1\kern-0.75\wd1\copy2}}

\newcommand{\pvint}{
\mathrel{\setbox1=\hbox{$\displaystyle\int$}\setbox2=\hbox{$-$}
\copy1\kern-0.90\wd1\copy2}}

\newcommand{\pvintab}{
\mathrel{\setbox1=\hbox{$\int_a^b$}\setbox2=\hbox{$-$}
\copy1\kern-1.15\wd1\copy2}}

\def\mpid{{M^2_\pi}}

\def\fpi{{F_\pi}}
\def\qbarq{{\bar u u}}
\def\mkbar{{\bar M_K}}

\def\mkd{{M^2_K}}
\setbox0=\hbox{$\scriptstyle=$}
\setbox1=\hbox{F}
\setbox2=\hbox{T}
\setbox3=\hbox{$\Sigma$}
\setbox4=\hbox{C}
\def\Fmat{\copy1\kern-\wd1\raise1.1\ht1\copy0}
\def\Tmat{\copy2\kern-\wd2\raise1.1\ht2\copy0}
\def\Sigmat{\copy3\kern-\wd3\raise1.1\ht3\copy0}
\def\Cmat{\copy4\kern-\wd4\raise1.1\ht4\copy0}
\def\MO{{Muskhelishvili-Omn\`es\ }}
\begin{document}
\vskip -1 truecm
\rightline{IPNO-DR 99-22}
%\rightline{\today}
\begin{center}
{\bf $N_f$ DEPENDENCE OF THE QUARK CONDENSATE}\\
{\bf FROM A CHIRAL SUM RULE}
\bigskip

Bachir Moussallam\footnote{Address after Sept. 1: MIT, Center for
Theoretical Physics, 77 Massachussets Av., Cambridge MA 02139-4307}

{\sl I.P.N., Groupe de Physique Th\'eorique}\\
{\sl Universit\'e Paris-Sud, F-91406 Orsay C\'edex}
\end{center}

\centerline{\large\bf Abstract}

How fast does the quark condensate in QCD-like theories
vary as a function of $N_f$ 
is inferred from real QCD using chiral perturbation theory at order one-loop. 
A sum rule is derived for the single relevant chiral
coupling-constant, $L_6$. A model independent lower bound is
obtained. 
The spectral function satisfies a Weinberg-type
superconvergence relation. It is discussed how this, together with
chiral constraints allows a  solid evaluation of $L_6$, based on
experimental $\pi\pi-K\bar K$ S-wave T-matrix input.
The resulting value of $L_6$ is compatible with a strong
$N_f$ dependence possibly suggestive of the proximity of a chiral
phase transition.

\bigskip

\noindent{\large\bf 1  Introduction}

By analogy with recent results obtained in supersymmetric
theories\cite{susy}, one expects that QCD-like theories will undergo a number
of phase transitions at zero temperature upon varying $N_f$, the number of
different flavour fermions, 
at fixed number of colours ($N_c=3$ in the following). If the
number of fermions is large $N_f>{11\over2} N_c$ the theory has no asymptotic
freedom and no confinement. Decreasing $N_f$ below ${11\over2} N_c$
one encounters a conformal phase as indicated by the fact that the
$\beta$-function at two loops has a zero\cite{zaks}.
Assuming the fermions to be all
massless, the chiral $SU(N_f)\times SU(N_f)$ symmetry
of the QCD action remains unbroken in this phase. 
If one further decreases $N_f$ to small values, $N_f=2-3$, then QCD
is in a confining phase in which the chiral group is spontaneously broken to
$SU(N_f)$. 
It is generally believed that, in this phase, the quark condensate is
non-vanishing and large\footnote{Experimental verification of this
conjecture necessitates specific and very precise data. This question 
is discussed in ref.\cite{knecht}.}. At larger $N_f$, there could exist
different phases of chiral symmetry breaking. 
A physical picture of such phases is proposed in ref.\cite{stern}. An
interesting open question concerns the value of $N_f^{crit}$ for which
a chiral phase transition takes place. Recent lattice results suggest that
a transition could occur for $N_f$ as small as four\cite{lattice}. In
an instanton vacuum model, the quark condensate ceases to be
non-vanishing for $N_f$ of the order of five\cite{instanton}, while
another theoretical model obtains a much larger value\cite{appelq}.

In this paper, we use the fact that nature solves ordinary QCD in order to
extract an information on how the quark condensate varies with $N_f$ for
small values of $N_f$. More specifically, if it can be shown that the ratio
\be
R_{32}={<\qbarq>_{N_f=3}\over <\qbarq>_{N_f=2}}
\en
is significantly smaller than one,
one may expect a rather small value of $N_f^{crit.}$.
In the functional integral, the dependence upon $N_f$ arises from the
fermion determinant part of the measure: setting all quark masses
equal, one has
\be
d\mu\equiv d\mu(A)\left( \det (i\Dslash+m)\right)^{N_f}\ .
\en
In other terms, it is a Dirac sea effect. In the quenched approximation,
which is often used in lattice simulations, the fermion determinant is set
equal to one and $R_{32}$ is exactly one. The same result also follows in the
leading large $N_c$ expansion of QCD, since the determinant contributes
to graphs with internal quark loops, which are subleading. 
Using chiral perturbation theory (CHPT), one can access the ratio 
\be\lbl{rtilde}
\tilde R_{32}= {<\qbarq>_{(m_u=m_d=m_s=0)}\over
<\qbarq>_{(m_u=m_d=0,m_s\ne0)}}\ .
\en
This ratio is different from $R_{32}$, but it is also a measure of the
influence of the fermion determinant in the evaluation of the quark
condensate. Again, this ratio would be exactly one in the leading large $N_c$
expansion or in the quenched approximation, for any value of the strange
quark mass $m_s$.
%%%%%lundi 12 juillet
The point here is that the
physical value of strange quark mass is sufficiently small compared to the
scale of the chiral expansion $\Lambda\sim1$ GeV, such that the chiral
expansion in $m_s/\Lambda$ makes sense and, at the same time, $m_s$
is not that small
so that $\tilde R_{32}$ will not trivially be close to one. In this paper,
we will provide an estimate of $\tilde R_{32}$.

The plan of the paper is as follows. In sec.2, the expression of
$\tilde R_{32}$ in CHPT at order one-loop is given. 
This expression involves a single low-energy coupling constant, 
$L_6(\mu)$, in the
nomenclature of Gasser and Leutwyler\cite{gl85}. In that paper, 
$L_6$ was simply assumed to be OZI suppressed. 
Here, we attempt a more careful
estimate on the basis of a chiral sum rule. Analogous chiral sum rules were
discussed in the recent literature\cite{dono} and eventually provide very good
precision\cite{luca}. In sec.3,  
a sum rule expression for $L_6$ in terms of the
correlation function $\Pi_6(s)$ of two scalar currents $\bar u u+\bar d d$
and $\bar s s$ is established and $Im \Pi_6(s)$ is shown to satisfy a
Weinberg-type sum rule. This will be an important constraint to our
evaluation. The construction of the spectral function is discussed in sec.4. 
Important ingredients are the pion and the kaon scalar form-factors which
can be related  to experimental information on pion-pion scattering using
analyticity, unitarity, high-energy constraints 
as well as low-energy constraints from chiral symmetry. 
This was first performed in ref.\cite{dgl}. Extension to the region of
1.5 GeV, where
important resonance contribution is expected is then discussed. Finally, the
result can be found in sec.6.

\noindent{\large\bf 2. Ratio of quark condensates from CHPT }

Consider QCD in a limit where $N_f$ quarks are exactly massless.
We will consider
the cases of $N_f=2$ and $N_f=3$ and assume, a priori, that
chiral symmetry is spontaneously broken in QCD in both cases and that the
value of the condensate is sufficiently large also in both cases such
that the
conventional chiral expansion \cite{gl84}\cite{gl85} applies.
In nature, none of the quark masses $m_u$, $m_d$
$m_s$ is exactly vanishing, but they are in an asymmetric configuration where
$m_u,\ m_d << m_s$ (by a factor of twenty or so) and $m_s$ is itself
sufficiently small compared to the scale of the chiral expansion $\Lambda$.
Using this fact, one can express the  
ratio $\tilde R_{32}$ as an expansion in powers of $m_s$.
At chiral order $O(p^4)$, making use of the formulae of
ref.\cite{gl85}, one obtains
\be\lbl{rel32}
\tilde R_{32}=1-{m_s B_0\over\fpi^2}
\left[
32L_6(\mu)-{1\over16\pi^2}\left(
{11\over9}\ln{m_s B_0\over\mu^2}+
{2\over9}\ln{4\over3}\right)\right]+O(m^2_s)\ ,
\en
At order $O(p^2)$  one has,
\be
m_s B_0={1\over2}\left(M^2_{K^+}+M^2_{K^0}-M^2_{\pi^+}\right),
\en
and this value can be used consistently in the equation above.
The size of $\tilde R_{32}$ depends on  the value of a single
low-energy coupling constant, $L_6$. 

The low-energy coupling constants of CHPT 
may be related to QCD correlation functions
evaluated near zero momenta. This can be exploited, in particular for
two-point functions, in order
to express these constants in the form of sum rules
using analyticity (and the fact that chiral correlators have non-singular
short distance behaviour). A number of these are
exhibited in ref.\cite{gl84}. A classic example 
concerns the coupling constant $L_{10}$ which is related
to the correlation function of two vector currents minus two axial currents.
A very reasonable estimate for $L_{10}$ can be obtained using simply
the idea of vector meson dominance as well as Weinberg sum rules\cite{eglpr}.
Our aim is to estimate $L_6$ along similar guidelines.

One specific reason for interest in $L_6$ is in connection with the
Kaplan-Manohar transformation\cite{KM}. These authors observed that the
effective lagrangian is left invariant under the following 
transformation of the quark
mass matrix,
\be\lbl{km}
{\cal M}\to {\cal M}+\alpha\left(32B_0\over F^2_\pi\right)
({\cal M^\dagger})^{-1}\det {\cal M},
\en
together with a transformation of certain low-energy constants. At chiral
order $O(p^4)$ three coupling constants are affected, $L_6$, $L_7$ and $L_8$, 
which get transformed as
\be
L_6\to L_6-\alpha,\quad L_7\to L_7-\alpha,\quad L_8\to L_8+2\alpha\ .
\en
One consequence is that using low-energy data alone, one can only determine
combinations which are invariant under this transformation and not the
individual values of $L_6$, $L_7$ and $L_8$. 
These values are of some importance. 
In particular, the value of $L_8$ determines the ratio of quark masses
$2m_s/(m_u+m_d)$ beyond the leading chiral order\cite{gl85}. It is therefore
of interest to explore means of separately determining 
these constants (or at least one of them).

\noindent{\large\bf 3. Sum rule for $L_6$ }

Consider the correlation function  of the two scalar, isoscalar currents
$\bar u u+\bar d d$ and $\bar s s$, 
\be
\Pi_6(p^2)=iB_0^{-2}\int d^4x e^{ipx} < T\left[ (\bar u u(x)+\bar d d(x))
\,\bar s s(0)\right] >_{c},
\en
where the subscript $c$ means that only connected graphs are to be retained.
The factor $B_0^{-2}$ is introduced to simplify forthcoming expressions and
furthermore makes $\Pi_6$ a renormalisation scale invariant object
and $B_0$ is defined as
\be
B_0=-\lim_{m_u=m_d=m_s=0}{<\bar u u>\over F^2_\pi}\ .
\en
For small
momenta we can express $\Pi_6(p^2)$ using CHPT. In particular, at zero
momentum, from CHPT at $O(p^4)$ one obtains,
\be\lbl{L6}
\Pi_6(0)=64 L_6(\mu) -{1\over16\pi^2}\left[ 2\ln{(m_s+m)B_0\over\mu^2}+
{4\over9}\ln {(4m_s+2m)B_0\over3\mu^2}+{22\over9}\right] + O(m,m_s).
\en
Here, and in the following, isospin breaking is neglected and we set 
$m_u=m_d=m$. 
This expression is at the basis of our sum rule estimate for $L_6$.
The quark condensate ratio $\tilde R_{32}$ has a very simple expression in
terms of $\Pi_6$. 
Combining equations \rf{rel32} and \rf{L6} one obtains
\be\lbl{rel32bis}
\tilde R_{32}=1-{\mkbar^2\over 32\pi^2\bar F^2_\pi}\left[
16\pi^2\bar\Pi_6(0)+{22\over9}\right] +O(m^2_s) ,
\en
where barred quantities are to be taken in the limit $m_u=m_d=0$. This
relation (and eq.\rf{rel32}) can be recovered alternatively by noting that
\be
{\partial\over \partial m_s}<\bar u u+\bar d d>=-B^2_0\Pi_6(0),
\en
and integrating this equation from $m_s=0$ to its physical value using
the CHPT expression \rf{L6}.

It is possible to derive a lower bound on $L_6$ and on $\bar\Pi_6(0)$ 
based on general properties of the QCD measure.
At first, it is not very difficult to show that $\Pi_6(0)$ must be positive
in the case of equal quark masses \cite{janpriv}. 
Let $Z$ be the partition function of euclidian QCD,
\be
Z=\int d\mu(A)\, e^{-S_{YM}(A)}{\rm det} (i\Dslash_A+M),\quad M=diag(m,m,m_s)\ .
\en
We can express $\Pi_6(0)$ in terms of $Z$ as
\be
\Pi_6(0)={1\over Z}{d^2 Z\over dm dm_s}-{1\over Z}{dZ\over dm}\,
{1\over Z}{dZ\over dm_s}\ .
\en
In the limit $m_s=m$ this can be written as an average of a manifestly positive
quantity,
\be
\lim_{m_s=m}\Pi_6(0)= <  \left( {\rm Tr}{1\over i\Dslash+m} -
<{\rm Tr}{1\over i\Dslash+m}>\right)^2 >\ ,
\en
where averages are defined as
\be
<O>={1\over Z}\int d\mu(A)\, e^{-S_{YM}(A)}{\rm det}(i\Dslash_A+M)
O(A)\ .
\en
This is because ${\rm Tr}\,{(i\Dslash+m)^{-1}}$ can be shown to be real
and the averaging is performed with respect to an integration
measure which is real and positive in euclidian QCD
(assuming the vacuum angle $\theta=0$
and a proper regularisation of the fermion determinant). 
%%%%% Mercredi 13 Octobre. %%%%
We can now apply this result on the positivity of $\Pi_6(0)$ in
conjunction with its one-loop expression, eq.\rf{L6}, setting all
three quark masses there equal to the physical $m_s$ value. One gets
\be
L_6(\mu)\ge {11\over4608\pi^2}
\left[\log{2m_s  B_0\over\mu^2}+1\right]+O(m_s)\ .
\en
Ignoring higher loop corrections, this gives
\be\lbl{bound}
10^3 L_6(M_\eta)\ge 0.35,\quad 16\pi^2\bar\Pi_6(0)\ge 1.57\ .
\en
This shows, in particular, that the condensate must be a decreasing
function of $N_f$.

Another property, which will prove an important constraint for the sum rule
estimate of $L_6$ is that $\bar \Pi_6$ satisfies a Weinberg-type 
sum rule (WSR)
\be\lbl{wsr}
\int_0^\infty Im\bar \Pi_6(s) ds = 0\ .
\en
The proof is analogous to that of the ordinary Weinberg sum
rules (see e.g. \cite{wbook} ). The operators in the operator-product
expansion at short distances must transform in the same way as
$(\bar u u+\bar d d)\,\bar s s$ under the chiral group. The masses $m_u$ and
$m_d$ being set equal to zero, the operator of lowest dimensionality
having the correct transformation property is $m_s(\bar u u+\bar d d)$ and it
has dimension four.
Furthermore, a factor of $(\alpha_s)^2$ is
generated from the fact that all connected graphs contain at least two gluon
lines. Taking into account the scale dependence of $\alpha_s$ and that of
$B_0$ in the perturbative region, we learn that $\bar \Pi_6$ vanishes
asymptotically faster than $1/q^2$,
\be\lbl{asy}
\bar \Pi_6(q^2)= {C <m_s (\bar u u + \bar d d)>\over
q^2 \left[\ln(-q^2)\right]^{2+24/27}}+O({1\over q^4}) ,
\en
with $C$ a constant, which implies the WSR eq.\rf{wsr}.
Furthermore, this behaviour ensures convergence of an unsubtracted
dispersion relation for $\bar \Pi_6$, so that one can express its value at
zero as
\be\lbl{drep}
\bar \Pi_6(0)={1\over\pi}\int_0^\infty ds\,{ Im\bar \Pi_6(s)\over s}\ .
\en
While the WSR must be
satisfied for arbitrary values of $m_s$, it is clear from eq.\rf{asy} that
convergence of the integral \rf{wsr} 
will be faster if $m_s=0$. In that situation,
one expects the sum rule to be saturated in an energy interval of, say, $[0-2]$
GeV, by analogy with the ordinary Weinberg sum rules.
Experimental data is known for $m_s\ne0$,
but while
there may be large differences {\sl locally} in $Im\Pi_6(s)$ upon
varying the value of $m_s$ (notably because of threshold effects), it is
expected  that these
differences will be smoothed out to a large extent in the integral. 
One therefore expects that
the WSR will be also approximately saturated in a finite energy region, for the
physical value of $m_s$.

Let us make some qualitative remarks on the practical significance of the
sum rule. In the large $N_c$ counting, $\Pi_6$ is suppressed compared to a
generic QCD correlation function, it is of order $O((N_c)^0)$ instead of
$O(N_c)$. As a byproduct, one observes that the contribution of a single
resonance is not enhanced by a factor of $N_c$
compared to the non-resonant background. In the large $N_c$ world,
the coupling of a resonance to either $\bar u u+\bar d d$ or $\bar s s$ will
be suppressed. For a glueball, both couplings will be suppressed. The real
world, in the scalar sector, seems to be quite different from these large $N_c$
considerations. For instance the $f_0(980)$ meson is found experimentally to
be rather light, narrow, and it couples strongly to both $K \bar K$
and $\pi\pi$ channels in violation to the large $N_c$ expectation. As a
consequence, one expects a strong contribution of the $f_0(980)$
to $Im \Pi_6$. In order to
satisfy the WSR \rf{wsr} the contribution from the
$f_0(980)$ has to be canceled by a higher energy contribution. It seems
plausible that this will be resonance dominated as well. The particle-data
book\cite{pdg} quotes several resonances in the 1.5 GeV region: the
f$_0$(1370), a rather wide resonance, the 
f$_0$(1500) which is well defined and rather narrow and
(possibly) the f$_0$(1700). A first guess is that the sum rule \rf{wsr} 
should be
essentially satisfied from an interplay between the f$_0$(980) and the
f$_0$(1500). 

The remaining problem is to estimate the couplings of these
resonances to the scalar currents. This cannot be extracted directly
from experiment because of the absence of a physical scalar 
iso-scalar source ( which
is the same reason why $L_6$, $L_7$, $L_8$ cannot be individually determined
from low-energy experiments\cite{leutw} ). One way of getting around this
difficulty, which was used in QCD sum rule estimates of the light quark masses
\cite{deraf1} is to impose smooth matching of the resonance contribution with
the low-energy domain, which is known from CHPT at leading order. 
This procedure can be checked to be a reasonable one in the case of vector
currents. 
Implementation of this idea in the present context is discussed below.

\noindent{\large\bf 4. Construction of the spectral function}

\noindent{\bf A. The role of two-body channels}

In the construction of $Im\Pi_6(s)$ it is convenient to consider separately
the two energy regions I) $0<s\lapprox 1\ {\rm GeV}^2$ and II)
$s\gapprox 1\ {\rm GeV}^2$.
Let us consider region I first. The only intermediate states allowed
to contribute to $Im\Pi_6(s)$ there are $2\pi$, $4\pi$ and $K\bar K$. When
$s<<1$ $\rm GeV^2$, the $4\pi$ 
contribution is suppressed by the chiral counting (being of
order $O(p^8)$ while the leading contribution is $O(p^4)$). Close to 1
$\rm GeV^2$, chiral counting is no longer effective, but it is found experimentally
that the $f_0(980)$ has very little coupling to $4\pi$
(in fact, no decay of the $f_0(980)$ into
four pions has been observed yet\cite{pdg}).  It is extremely likely, then,
that the $4\pi$ contribution to $Im\Pi_6(s)$ is negligible 
in this whole energy range. As a result, the spectral function can be
expressed in terms of the pion and of the kaon scalar form-factors. It
is convenient to introduce the following normalisations,
\bea\lbl{F1}
&&F_1(s)={1\over B_0}\sqrt{3\over2}<0\vert \bar uu+\bar dd\vert\pi^0\pi^0>\quad 
G_1(s)={1\over B_0}\sqrt{3\over2}<0\vert \bar ss\vert\pi^0\pi^0>\nonumber\\
&&F_2(s)={1\over B_0}\sqrt{2}<0\vert \bar uu+\bar dd\vert K^+K^->\quad 
G_2(s)={1\over B_0}\sqrt{2}<0\vert \bar ss\vert K^+K^->\ .
\ena
The values of these form-factors
at $s=0$ are proportional to the derivatives of $M^2_\pi$ and
$M^2_K$ with respect to the quark masses. At leading chiral 
order, one has
\be\lbl{fchir}
F_1(0)=\sqrt{6}\qquad G_1(0)=   0    \qquad
F_2(0)=\sqrt{2}\qquad G_2(0)=\sqrt{2}\ .
\en
One-loop corrections to these values can consistently be ignored because
they are of the same order as the $O(p^6)$ contributions in eq.\rf{L6}.
In the energy range I, the spectral function has the following expression
\be\lbl{specff}
Im \Pi_6(s)={1\over16\pi}\sum_{i=1}^2\sqrt{s-4M^2_i\over s}\,F_i(s)G_i^*(s)
\theta(s-4M^2_i)\ ,
\en
(where $M_1\equiv M_\pi$, $M_2\equiv M_K$). 

We consider now the energy region II. More approximations will have to be
made in this region. We will work out the spectral function from
several models in order to illustrate how the WSR can be
satisfied. For the final purpose of evaluating the dispersive integral
eq.\rf{drep} we will mainly rely on information from energy range I.
As $s$ increases a new two-body channel opens, $\eta\eta$. At some point,
the $4\pi$ channel will become important. Studies of $\pi\pi$ scattering
suggest that this should happen at $\sqrt{s}\gapprox 1.4$ GeV. This can be
seen from the $\pi\pi$ inelasticity: below 1.4 GeV, inelasticity is found to
be saturated to a good approximation by a single inelastic channel, $K\bar
K$  (see e.g. fig. 7 of ref.\cite{au}) and then one
observes a strong onset of $\pi\pi\to 4\pi$. It is very likely that this
is caused by the presence of 
the nearby scalar resonances $f_0(1370)$ and $f_0(1500)$ which
were both observed to couple to four pion states
\cite{adamo},\cite{amsler94},\cite{abele96}. 
Another experimental finding of these
references is that the $4\pi$ system, in this energy region, likes to
cluster into two resonances. This suggests that in an energy range
sufficiently large to saturate the chiral sum rule, the contributions to the
spectral function are either two-body channels or behave to a good
approximation as quasi two-body channels. We will utilise below a
model\cite{kll1} in which the $4\pi$ system is treated as an effective
$\sigma\sigma$ two-body channel. Correspondingly, we will introduce the
scalar form-factors,
\be
F_3(s)={1\over B_0}<0\vert \bar uu+\bar dd\vert \sigma\sigma>\quad
G_3(s)={1\over B_0}<0\vert \bar ss\vert \sigma\sigma>\ .
\en
This is certainly somewhat schematic as it is known that $4\pi$ also
clusters as an effective $\rho\rho$ channel. In this model, furthermore, the
$\eta\eta$ channel is ignored. This is a questionable approximation,
perhaps, although there is experimental evidence that the coupling
of $\eta\eta$ to $\pi\pi$
appears to be relatively suppressed\cite{alde}\cite{longacre}. 
In the quasi two-body approximation of
multimeson channels one can express the spectral function in terms of
form-factors $F_1(s),\ldots ,F_n(s)$ and $G_1(s),\ldots ,G_n(s)$ in the same
way as eq.\rf{specff} except that the sum extends from 1 to n. 
Upon introducing effective two-body channels, one faces the difficulty
that one can no longer rely on CHPT in order to determine the values of the
form-factors at the origin. These values are needed 
in the construction to be described below. 
In practice, we will make the simple ansatz
that these values are vanishing,
\be
F_i(0)=0,\quad G_i(0)=0,\ i\ge3\ .
\en
Evaluation of the scalar form-factors of the
pion and the kaon was discussed in ref.\cite{dgl}, based on a set of
\MO equations. We review this evaluation and its extension below.

\noindent{\bf B. \MO representation of scalar form-factors}

The  form-factors $F_i(s)$ which occur in the expression for the spectral
function \rf{specff} (generalised to $n$ channels) 
are themselves analytic functions everywhere in the
complex plane except for a right-hand cut. Let $T_{ij}$ be the T-matrix
elements which describe scattering among the various channels. A standard
normalisation is adopted where the S and T matrices are related as
\be
S_{ij}=\delta_{ij}+2{\rm i}\sigma_i^{1\over2} T_{ij} \sigma_j^{1\over2}
\theta(s-4M_i^2)\theta(s-4M_j^2),\ {\rm with}
\ \sigma_i=\sqrt{s-4M_i^2\over s}\ .
\en
The discontinuity of the form-factors along the cut,   
generated from the two-body channels considered above, has the following form,
\be\lbl{disc}
Im F_i(s)=\sum_{j=1}^n T^*_{ij}(s) \sigma_j(s) F_j(s) \theta(s-4M_j^2)\ .
\en
We expect the form-factors to vanish asymptotically as\cite{brodsky}
\be
F_i(s)\sim 1/s,\ s\to\infty,
\en
and therefore to satisfy an unsubtracted dispersion relation.
Clearly, the approximation of quasi two-body channels cannot hold 
for arbitrarily large energies and 
eq.\rf{disc} is a reasonable  approximation to the exact
discontinuity only in a finite energy range.
However, as we are interested in constructing 
$F_i(s)$ in a finite
energy region also, say below two GeV, the detailed behaviour of the spectral
function at much higher energies is unimportant and we may as well assume
that eq.\rf{disc} holds up to infinite energies, only requiring that the
T-matrix behaves in a way that ensures the correct asymptotic decrease of
the form-factors. Under these assumptions the form-factors must satisfy a
set of coupled \MO \cite{muskh}\cite{omnes}(MO for short) 
singular integral equations,
\be\lbl{moeq}
F_i(s)={1\over\pi}\sum_{j=1}^n
\int_{4M^2_\pi}^\infty ds' {1\over s'-s}\, T^*_{ij}(s')
\sqrt{s'-4M^2_j\over s'}\,\theta(s'-4M^2_j) F_j(s')\ .
\en
One observes that
off-diagonal T-matrix elements are needed outside of the
physical scattering region. Except in the one-channel case, this means that
one not only needs physical scattering data but also a parametrisation model
which allows for extrapolation.

\noindent{\bf C. Asymptotic conditions on the T-matrix}

Let us now specify which asymptotic conditions
are required from the T-matrix. Consider first the single channel case 
for which
an analytic solution to the MO equation is available\cite{muskh}\cite{omnes},
\be
F(s)=P(s)\Omega(s),\quad
\Omega(s)=\exp\left[{s\over\pi} \int_{4M^2_\pi}^\infty ds'{1\over
(s'-s)s'}\,\delta(s')\right] ,
\en
where $\delta(s)$ is the scattering phase-shift and $P(s)$ an arbitrary
polynomial. Integrating by parts, it is simple to verify that, as
$s\to\infty$ one has
\be
\Omega(s)\to s^{-l},\ l={1\over\pi}(\delta(\infty)-\delta(4M^2_\pi))\ .
\en
Compatibility with the assumed high-energy behaviour of the form-factor is
ensured provided $l\ge 1$. It is not difficult to see how this
condition extends to the situation of $n$ coupled channels\cite{muskh},
even though no
analytical solution is known in general. Let us form a vector $\vec F$ of
components $(F_1(s),\ldots ,F_n(s))$; we learn from Muskhelishvili's book that
there will be in general $n$ independent solution vectors
$\vec F_a$, $a=1,\ldots n$ to the
set of equations. Let us form an $n\times n$ matrix from these 
\be
\Fmat(s)\equiv\left( \vec F_1(s),\ldots ,\vec F_n(s)\right)\ .
\en
All matrix elements of $\Fmat(s)$ are analytic functions of $s$ in the cut
complex plane and the discontinuity across the cut can be formulated in
matrix form,
\be
\Fmat(s+i\epsilon)=(1+2i T\,\Sigma)\,\Fmat(s-i\epsilon),\quad
\Sigma_{\,ij}=\delta_{ij}\sigma_i(s)\theta(s-4M^2_i)\ .
\en
Taking the determinant of both sides, we obtain a one-dimensional
discontinuity equation,
\be
f(s+i\epsilon)=D(s) f(s-i\epsilon),\quad f=\det\Fmat .
\en
As $f(s)$ is also an analytic function, this equation can be recast as
a one-channel MO equation. As a consequence, the determinant of the solution
matrix $\Fmat$ can always be expressed in analytical form even though the
individual entries are not known analytically. This is an interesting
property which we have used as a check of the accuracy of 
our numerical calculations.
It is easy to verify that, for a given value of
the energy $s$ with $m\le n$ channels being open, 
$D(s)$ is the determinant of the
$m\times m$ S-matrix so that it is a complex number of unit modulus,
\be
D(s)\equiv\exp(2i\Delta(s))\ .
\en
Letting $s$ go to infinity, $\det\Fmat$ behaves as
an inverse power of $s$, so that the matrix $\Fmat$ must be of the following
form,
\be
\lim_{s\to\infty}\Fmat(s)={1\over s^\nu}\,\Cmat\ ,
\en
with $\Cmat$ a constant $n\times n$ matrix with non-vanishing determinant. 
Taking the determinant of this equation 
and using the one-channel result one finds that
$1/s$
asymptotic behaviour is ensured by the asymptotic condition,
\be\lbl{asymp}
\Delta(\infty)-\Delta(4M^2_\pi)\ge n\pi\ .
\en
For instance, in the case of three channels, 
the T-matrix must be such that the sum of the 
three eigen-phase-shifts sum up to $3\pi$ (or more)
when the energy goes to infinity. If the sum is exactly $3\pi$, then the
form-factors are uniquely determined at any energy from their values at zero.

\noindent{\bf D. Models of $\pi\pi$ scattering T-matrix}

In principle, $\pi\pi$ phase-shifts and inelasticities
can be determined from di-pion production experiments,
in which high-energy pions are scattered 
on proton targets (e.g. \cite{MMSbook}). A
major source of information in this area so far, is 
from the high statistics experiment
by the CERN-Munich collaboration\cite{cernmu} from which $\pi\pi$ S-matrix
elements were extracted by a number of people\cite{MMSbook}. Various
determinations of S-wave phase-shifts
are generally in reasonable mutual agreement below 1.4 GeV while
marked differences are seen above.
From these early analysis there was  no clear evidence for resonances at
1.4 or 1.5 GeV. The CERN-Munich data themselves, however, are not
incompatible with the presence of scalar resonances at these energies.
This was demonstrated recently by Bugg et al.\cite{bugg} who obtained a 
good fit to the CERN-Munich data while constraining the S-matrix to have
resonance poles and residues conforming to the PDG results.
Unfortunately, it is not possible to use their
parametrisation of the S-matrix for solving the MO equations 
because, on the one hand, it is not designed to satisfy the full set of
two-body unitarity constraints and, moreover, the corresponding
T-matrix parametrisation does not allow for extrapolation away 
from the physical scattering region.
A  set of $\pi\pi$ S-wave
scattering phase-shifts and inelasticities was obtained recently\cite{klr},
based on high statistics di-pion production data employing {\sl polarised} 
proton
targets\cite{cerncrac}. In principle, polarisation information is extremely
useful in reducing the problems of phase ambiguities. 
Two solutions consistent with unitarity were found,
called up-flat and down-flat. We will consider the
latter one only here, because on the one hand, 
it is in good agreement with earlier
phase-shift determinations below 1.4 GeV and on the other hand, 
up-type solutions
can usually be eliminated upon using the Roy equations\cite{roy}\cite{MMSbook}
which encode crossing symmetry and high-energy constraints.
This determination shows a marked
resonance effect in the 1.4-1.5 GeV region and thus appears as a good
candidate for use in our sum rule analysis. One notes that, above 1.4 GeV, 
the $\pi\pi$  phase-shifts determined by \cite{klr} and those determined in
ref.\cite{bugg} are not in good agreement. 

For the purpose of solving the
MO equation system one further needs a T-matrix parametrisation allowing
convenient (and reliable) extrapolation below physical thresholds. As pointed
out in ref.\cite{dgl} a useful check on the extrapolation of $T_{12}$
is to compare it to the chiral expansion in the region where the latter is
valid. Close to $s=0$ one has,
\be
T_{12}={\sqrt{3}\over64\pi F^2_\pi}s + O(p^4)\ .
\en
One-loop corrections to this result
have been worked out\cite{BKM}\cite{roessl}. 
%%%%The TW  model %%%%%
A simple T-matrix model, which is very useful for performing checks of
numerical calculations is that proposed by Truong and Willey\cite{TW}. 
This model has the property that the OM set of equations can be solved
analytically.
%%% The KLL model %%%%%
A somewhat more sophisticated  T-matrix model, fitted to
reproduce the $\pi\pi$ data of ref.\cite{klr} was proposed in 
ref.\cite{kll1}. Fits with both 2-coupled  and 
3-coupled channels were performed. 
In this model, unitarity is ensured by solving a
Lippman-Schwinger equation with a potential matrix 
chosen to  have the following
separable form,
\be\lbl{pot}
V_{ij}(p,q)=\sum_{l,m}{1\over p^2+\mu^2_{il}}{1\over
q^2+\mu^2_{jm}}\lambda_{lm}\ .
\en
The T-matrix can be computed analytically and it can be checked to have the
correct chiral magnitude at low-energy (in other terms it vanishes linearly
with $s$ and $M^2_\pi$). 
It seems possible to adjust the
parameters (and also the propagator) in order to reproduce exactly the
correct T-matrix chiral expansion at $O(p^2)$
and even, we believe, at $O(p^4)$, but this has not yet been done. 
In this model, the OM equations must be solved numerically. 
For this purpose, we
have developed an algorithm which is described in the appendix.

The $\lambda$ and $\mu$
arrays in eq.\rf{pot} are constant parameters fitted to the data.
In the case of 3-coupled channels, the available data is not
sufficiently constraining and several different sets of parameters can
provide comparable fits. 
Two different sets of parameters were obtained in refs.\cite{kll1} (called A
and B) and two further sets in ref.\cite{kll2} (called E and F). Sets A, B
and E generate fits with comparable $\chi^2$ with the set of data considered
in ref.\cite{kll1}. Set F has a good $\chi^2$ at low energy only. Close to
1.4 GeV it has a very narrow resonance, which, perhaps, could be interpreted
as a glueball. Although not producing a very good $\chi^2$, the authors of
ref.\cite{kll2} suggest that this scenario is not totally excluded by the data.
The data which was used in these fits
consists in a) The set of $\pi\pi$ phase shifts $\delta_\pi(E)$
and inelasticities $\eta_\pi(E)$
as determined in ref.\cite{klr} in the energy range
$ 0.6\le E_{\pi\pi} \le 1.6$ GeV 
and b)the set of phases $\phi_{12}(E)$ of the $\pi\pi\to K\bar K$ amplitudes
from the particular experiment of Cohen et
al.\cite{cohen}. One must keep in mind here that there is some discrepancy
in the lower energy part
between the result of this experiment and others, notably by Etkin et
al.\cite{etkin} as far as the phase is concerned.
This point is discussed in
some detail by Au et al.\cite{au}, whose K-matrix parametrisation could more
easily reproduce the latter phase results.
Regrettably,
the absolute values of the $\pi\pi\to K\bar K$ amplitudes, which are also
available from experiment, were not included in the fits of 
refs.\cite{kll1}\cite{kll2}.
The various parameter sets differ in the behaviour of the phase-shift in
the third channel, $\delta_3(E)$, which is unconstrained by experiment and
also, to some extent, on the detailed structure of inelasticities. These
differences, as we will see, will result in fairly different behaviour of
the spectral functions as well, so that the sum rule \rf{wsr} appears as an
interesting theoretical constraint in this kind of analysis.

The T-matrices generated from
this model do not satisfy the asymptotic
constraints eq.\rf{asymp} neither in the 2-channel case nor
(for any of the parameter sets discussed above) in the 3-channel case.
Thus, they cannot be used up to infinite energies for our purposes.
We must  impose the proper asymptotic behaviour, i.e. that the
eigen-phase shifts must sum to $2\pi$ in the case of two channels and
$3\pi$ for three channels, by hand\footnote{These are the minimal
asymptotic values which ensure existence of a solution. We will
assume that a possible further raise above the minimal values can only
occur for $\sqrt{s}\gapprox 2$ GeV and will have no influence on lower
energy results.}. 
For this purpose,
we have introduced a cutoff energy $E_0$. For $E\le E_0$
the T-matrix is computed from the model, while for $E>E_0$
the phase-shifts are interpolated as follows
\be\lbl{cutoff}
\delta_\pi(E)=n\pi+(\delta_\pi(E_0)-n\pi) f\left({E\over E_0}\right),\quad
\delta_i(E)=\delta_i(E_0) f\left({E\over E_0}\right),\ i\ge2,
\en
with $n$ the number of channels and the cutoff function $f(x)=2/(1+x^m)$. 
In practice, we have taken $E_0=1.5$ GeV which insures a smooth raise of
$\delta_\pi(E)$ and $m=3$. Changing these parameters will modify the
details of the shape of the spectral function in the higher energy
region. 
Inelasticities are computed from the model
in the whole energy range. 
Other elements of the S-matrix can then be deduced 
from unitarity and continuity. 

\noindent{\large\bf 6. Results}

\noindent{\bf A. Two-channel models}

We first calculate the 
scalar form-factors and the spectral function from $\pi\pi-K\bar K$ 
2-channel models. The result for the spectral function, using 
the T-matrix model of
Au et al.\footnote{As observed in ref.\cite{dgl} upon using the values of
the parameters at the precision given in the Au et al. paper,
a spurious very narrow
resonance appears close to the $K \bar K$ threshold: we removed this
resonance by linearly interpolating the $\pi\pi$ phase-shift on both sides.
If not removed, the spectral function would be identical to that shown in the
figure except at the very position of the resonance where a very narrow dip
would appear.
}
\cite{au} is shown in Fig.1, together with the result from the 2-channel
version of the potential model of ref.\cite{kll1}. 
Consider first the region $\sqrt{s}\le 1$ GeV: there, the
spectral functions from the two models have the same sign and are 
rather similar in shape. 
For sum rule applications it is useful to introduce the following 
spectral function integrals in this energy region
\be\lbl{I_i}
I_n=16\pi\int_{4M^2_\pi}^{4M^2_K} {Im \Pi_6(s)\over s^n}\,ds\ .
\en
Some numerical 
results for the integrals $I_n$, $n=0,1$ are shown in table 1 below. The
predictions from the two models are seen to differ by less than 10\% 
for these quantities.

%------------Table 1--------------------------%
\begin{table}[hbt]
\begin{center}
\begin{tabular}{|c|c|c|c|c|c|}\hline
\      &Au    & KLL  & set A& set E& set F \\ \hline
$I_0$  & 3.30 & 3.09 & 4.18 & 3.38 & 2.69  \\
$I_1$  & 4.92 & 4.73 & 6.56 & 5.74 & 4.04  \\ \hline
\end{tabular}
\end{center}
\caption{\sl Values of the integrals $I_0$ and $I_1$ (see eq.\rf{I_i}) from
the 2-channel T-matrices labeled as Au\cite{au} and KLL\cite{kll1}. The
last three columns correspond
to different parameter sets of the 3-channel T-matrix model of  Kaminski et
al.\cite{kll2} }.
\label{Table 1}
\end{table}
%---------------------------------------------%

We have also computed, for these two models, 
the low-energy observables associated with pion form-factors. 
In the neighbourhood of $s=0$, one defines (we follow the notations of
ref.\cite{dgl})
\be
F_1(s)=F_1(0)\left[ 1+{1\over6}<r^2>^\pi_s\,s+c_\pi s^2+\ldots \right] ,
\en
and, similarly, for the matrix element of the $\bar s s$ current
\be
\sqrt{2\over3}\bar M^2_K G_1(s)=d_F s\left[ 1 +b_\Delta s+\ldots \right].
\en
As shown in ref.\cite{dgl} the parameter $d_F$ is proportional to the
derivative of $F_\pi$ with respect to the strange quark mass. Upon using the
T-matrix from Au et al., we have verified that our calculation reproduces
the results obtained previously\cite{dgl}\cite{gm}. The numbers are
displayed in table 2. We also show the results corresponding to
the T-matrix from ref.\cite{kll1}. 
The numbers quoted in the table correspond to an
improved T-matrix where the $\pi\pi$ phase-shift is constrained in the
low-energy region $\sqrt{s}\le 0.6$ GeV in order to match the predictions from CHPT
at two-loops for the scattering length and the scattering
range\cite{nous}\cite{colangelo} i.e.
$a_0^0=0.21$, $b_0^0=0.26 M^{-2}_{\pi^+}$. If we do not make this
modification, the results from the T-matrix of ref.\cite{kll1} would be 
only slightly different. For instance, one
would have, $<r^2>^\pi_s=0.580\ fm^2$, $d_F=0.075\ \rm{GeV^{-2}}$,
reflecting the reasonable low-energy behaviour of the T-matrix in this
particular model. These
results are compatible with the known low-energy coupling constants from
CHPT. At order one loop, the chiral expansion of $<r^2>^\pi_s$ and
$d_F$ involve the constants $L_4$ and $L_5$
\bea\lbl{L4L5}
&& <r^2>^\pi_s={24\over F_\pi^2}\left\{
2L_4(\mu)+L_5(\mu)-
{1\over32\pi^2}\left[ \log{\mpid\over\mu^2}+{1\over4}
\log{\mkd\over\mu^2}+{4\over3}\right]\right\}\nonumber\\
&& d_F={8\mkbar^2\over F_\pi^2}\left\{ L_4(\mu)-{1\over256\pi^2}\left[
1+\log{\mkbar^2\over\mu^2}\right] \right\}\ .\\
\ena
While $L_4$ is not easily determined elsewhere, 
$L_5$ can be extracted from the ratio
of $F_K/F_\pi$ and this gives\cite{gl85} $L_5(M_\rho)=1.4\pm0.5\,10^{-3}$.
Using eqs.\rf{L4L5} and the results from Table 2, one deduces
\be\lbl{l4l5}
L_4(M_\rho)\simeq 0.4\,10^{-3},\quad L_5(M_\rho)\simeq1\,10^{-3}\ .
\en
The value of $L_4$ can be considered as a prediction, and the result for
$L_5$ appears to be compatible with  $F_K/F_\pi$. One must
note, though, that this agreement might be somewhat fortuitous because
the error of this determination is rather large. If we assume, for instance,
10\% relative errors on $<r^2>^\pi_s$ and on $d_F$, the resulting
uncertainty on $L_5$ would be $\Delta L_5(M_\rho)=\pm 0.7\,10^{-3}$.

%---------------Table 2-----------------%
\begin{table}[hbt]
\begin{center}
\begin{tabular}{|c|c|c|c|c|c|c|}\hline
\  &$<r^2>^\pi_s$ & c$_\pi$ & d$_F$ & b$_\Delta$ &$10^3\, L_4$&
$10^3\, L_5$\\ \hline
Au   &0.585 & 10.50 & 0.087 & 3.29 & 0.41 & 0.81 \\
KLL  &0.605 & 10.81 & 0.076 & 3.45 & 0.36 & 1.09 \\ \hline
set A&0.609 & 10.86 & 0.134 & 3.12 & 0.61 & 0.62 \\
set E&0.583 & 10.58 & 0.144 & 3.15 & 0.66 & 0.29 \\
set F&0.653 & 11.51 & 0.047 & 3.80 & 0.23 & 1.79 \\ \hline
\end{tabular}
\end{center}
\caption{\sl Results for some low-energy observables and corresponding
chiral coupling constants at $O(p^4)$.
$<r^2>_s^\pi$ is in $fm^{-2}$ and other quantities in appropriate 
powers of GeV.
The first two lines correspond to 2-channel T-matrices and the the last
three lines to 3-channel models. The labeling is the same as in Table 1.}
\label{ Table 2}
\end{table}
%---------------Table 2-----------------%

In the region $E>1$ GeV now, the spectral functions from the two models
differ considerably, see Fig.1. That corresponding to the model of ref.\cite{kll1} 
exhibits a strong
resonance effect. Its contribution to the integral goes in the sense of
canceling the positive contribution from the $f_0(980)$. This is
qualitatively as expected from the WSR \rf{wsr} and indicates that it seems
possible to satisfy this constraint  in a 2-channel model.
Quantitatively, however, if one uses the set of parameters of ref.\cite{kll1}
without alteration,
one finds that the contribution from the $f_0(1500)$ is 
somewhat too strong and
overcompensates that of the $f_0(980)$. At any rate, in the 1.5 GeV region
it is no longer a good approximation to retain only two channels in
unitarity relations. Other two-body channels are open like $\eta\eta$,
and experiment indicates significant coupling of the $f_0(1500)$
to the $4\pi$ channel as well. We will now investigate how such additional
channels can affect the results in an approximation of a single effective
additional channel.

\noindent{\bf B. Three-channel models}

We have computed the spectral function based on the
3-channel T-matrix model
from ref.\cite{kll1} using several parameter sets determined in this
reference and in a subsequent one\cite{kll2}.
Results are shown in Figs.2-4. Fig.2 corresponds
to the parameter set A from
ref.\cite{kll1} (we recall that set B from this reference was discarded
because $T_{12}$ has an unphysical low-energy pole in this case)
and figs.3,4 correspond to the sets E and F from ref.\cite{kll2}
respectively.
Again, let us consider first the energy region $\sqrt{s}\le 1$ GeV: there,
the spectral functions  from the 3-channel models
are comparable to those from the 2-channel ones. This can be seen, for
instance, for the integrals $I_0$ and $I_1$ displayed in table 1. Also one
can see from Table 2 that the result for $<r^2>^\pi_s$ are rather stable. A
somewhat less stable quantity is $d_F$, the derivative of
the pion form-factor of the $\bar s s$ current, which increases in the 
3-channel model for parameter sets A and E. This results in a slight
increase of $L_4$ and a significant
decrease of $L_5$ which becomes too small 
and incompatible with $F_K/F_\pi$. 
On the contrary, a large value of $L_5$ emerges if one uses parameter
set F. Keeping in mind the uncertainty in the determination of $L_5$
from the scalar form-factors, none of the 3-channel parameter sets
considered here is as satisfactory as the 2-channel models as far as
the very low energy behaviour of the form-factors is concerned.

Let us now consider the energy region  $E > 1$ GeV.
One observes from Figs.2-4 that a variety of shapes can
get generated from different parameter sets.
Set F (see Fig.4 ) has a negatively
contributing resonance, but it is much too strong and does not obey
the WSR eq.\rf{wsr}. Set A has a positively
contributing resonance and does not obey the WSR constraint
either. Set E (see Fig.3)
displays a more complicated structure:
the $f_0(1500)$ has also a positive contribution but there is a wide
negative contribution centered at 1.7 GeV, and the WSR
is approximately obeyed.

\noindent{\bf C. Estimate of $\bar\Pi_6(0)$}

The main  conclusion from the above results is that while there seems to be
reasonable agreement on the shape of the spectral function in the energy
range $\sqrt{s}\le 1$ GeV, its structure above 1 GeV is subject to considerable
uncertainty. In models with more than two coupled channels the
parameters are not sufficiently constrained from the experimental
$\pi\pi$ and $K\bar K$ data. Another source of uncertainty in those
models which, again, can be checked to affect the spectral function
above one GeV concerns the values of the form-factors at the origin
$F_i(0),\ G_i(0),\ i\ge 3$ which are not given from chiral symmetry. 
At least, we have seen that there exist models which fit the data and can
also accomodate the WSR constraint. 

In order to calculate $\bar\Pi_6(0)$ from 
the spectral representation eq.\rf{drep} one
needs, in principle, to know the spectral function both below and above 1
GeV. However, the lower energy range is expected to generate the largest
contribution. Qualitative information on the high energy sector, such
as the existence of the WSR constraint and the experimental position
of the resonances is sufficient if one is not asking for a very high
precision. 
Firstly, we expect the contribution to $\bar\Pi_6(0)$ from
the range $\sqrt{s}>1$ GeV to be negative, giving the upper  bound,
\be\lbl{ubound}
\bar\Pi_6(0)\lapprox{1\over16\pi^2}I_1\ .
\en
The most plausible scenario is that of 
a single resonance dominated contribution around $\sqrt{s}\simeq 1.5$
GeV. In this
scenario, the following estimate of $\bar\Pi_6(0)$ is valid, 
\be\lbl{simple}
\bar\Pi_6(0)\simeq {1\over16\pi^2}\left(I_1 - {I_0\over(1.5)^2}\right)\ ,
\en
in which the WSR has been used. In this case, the correction from the
higher energy range is approximately 30\%.
The other possibility is that several resonances, the
$f_0(1500)$ and the $f_0(1700)$, are playing a role in the sum rules. 
If the two resonances make negative contributions one expects that the
correction to $\bar\Pi_6(0)$ will be smaller than 30\% because the
$1/s$ factor suppresses the $f_0(1700)$ contribution.
If the contribution from the $f_0(1500)$ is positive the correction is
even smaller (this was realised in one of the 3-channel models
considered above).
The last possibility is that of a negative $f_0(1500)$ and a positive 
$f_0(1700)$ in which case the contribution from the higher
energy region will be
largest, but simple estimates like \rf{simple} show 
that a 50\% correction is a generous upper limit. This gives us a
lower bound on $\bar\Pi_6(0)$,
\be
\bar\Pi_6(0)\gapprox {1\over32\pi^2} I_1\ .
\en
From these
considerations and the numbers of table 1 we infer  that 
the value of $\bar\Pi_6(0)$ must lie in the following range
\be
2\lapprox 16\pi^2\bar\Pi_6(0)\lapprox 6\ .
\en
Using chiral perturbation theory to one loop, eq.\rf{L6}, 
this result can be recast into
an estimate of the coupling constant $L_6$
\be\lbl{L6sr}
0.4\lapprox  10^3 L_6(M_\eta)\lapprox 0.8\ .
\en
We note that the bound \rf{bound} obtained in sec.3 is satisfied. 
This number can be compared with the estimate of ref.\cite{gl85}
\be\lbl{L6gl}
L_6(M_\eta)=(0.0\pm0.3)\,10^{-3}\ .
\en
%%%%jeudi 14
The central value there, is obtained from the assumption that the OZI
rule applies. Indeed, the OZI rule implies that $\tilde R_{32}$ is
identically equal to one, and inserting $L_6(M_\eta)=0$ in eq.\rf{rel32} 
one finds $\tilde R_{32}=0.96$ which is very close to one.

The value of  $L_6$ that we obtained from the sum rule
implies the following result for the ratio of  quark condensates 
$\tilde R_{32}$
\be
\tilde R_{32}\simeq 1-0.54\pm0.27\ .
\en
In order to obtain this estimate, 
we have used for $L_6$ the central value which
emerges from the sum rule discussion above and, for the error, we 
have used the same
value as that estimated in ref.\cite{gl85} (see eq.\rf{L6gl} ). Within the
substantial error band, the main observation is that the deviation
of the quark condensate ratio from one is negative, and it seems to be
rather large.

\noindent{\large\bf 7. Summary}

We started by noting that the sensitivity of the quark condensate
on $N_f$ can be tested by studying its variation as a
function of the strange quark mass. This variation may be related to the
correlation function $\Pi_6(q^2)$. Two
different expressions of $\Pi_6(0)$ are equalled, one based on chiral
perturbation theory and one which uses a dispersive representation. 
We discussed the spectral function which enters this
dispersive integral. We first argued that $Im \Pi_6$ satisfies a
Weinberg-type sum rule. This sum rule essentially relates resonance
contributions from the two energy regions, $0\le s \le 1$ ${\rm GeV}^2$  and
$1\le s \lapprox 4$ ${\rm GeV}^2$. In the first energy region the spectral
function can be expressed with good accuracy in terms of scalar form-factors
of the pion and the kaon. In turn, these form-factors can be constructed
from experimental scattering data on $\pi\pi$ and $K\bar K$ following the
method of ref.\cite{dgl}. The determination of the spectral function in the
higher energy range is more uncertain. We considered the prediction from a
model which treats the $4\pi$ channel as an effective two-body channel. The
influence of including this third channel in unitarity relations was found
to have relatively minor influence on results in the lower energy region but
has a strong influence on the region above one GeV. The dispersive
integral, fortunately, receives its main contribution 
from the lower
energy range. Using experimental information on the position of the
resonances, as well as the WSR constraint, allowed us to obtain an
estimate of $\bar\Pi_6(0)$.

The conclusion of this analysis is that the properties of the $f_0(980)$
meson translate into a value of the coupling constant $L_6$ which is
significantly different from that expected from the OZI rule (or,
alternatively, from large $N_c$ considerations).
If one uses the central value obtained for $L_6$, one finds that the
condensate $\qbarq$ decreases by as much as a factor of two as one decreases
the mass of the strange quark mass from its physical value down to zero. A
qualitatively similar behaviour is expected  if one varies $N_f$ from
$N_f=2$ to $N_f=3$. 
This surprising result is possibly suggesting that $N_f=3$ is not extremely far
from a chiral phase transition point.
One must bear in mind, however, that the relationship
that we used between $\Pi_6(0)$ and the condensate ratio receives
corrections from two-loop CHPT. It remains to be seen whether these are
significant or not.

\noindent{\large\bf Acknowledgments}
Robert Kaminski is thanked for discussions and communicating several data
files. Jan Stern is thanked for discussions, suggestions and comments on the
manuscript. 
This work is partly supported by the EEC-TMR contract ERBFMRXCT98-0169.

\smallskip
\noindent{\large\bf Appendix:} {\bf Numerical method}

The general idea for solving a linear integral equation
is to approximate it by an ordinary linear system of equations
by discretising the integral. The main
difficulty in the case of the MO equation is to handle the
principal-value integral with high accuracy.
Let us illustrate the method we have used 
on the one-channel MO  equation, the generalisation to several channels
is straightforward. 
First, one can transform the equation into one for the real part of the
form-factor, $R(s)=Re( F(s) )$,
\be\lbl{onedim}
R(s)={1\over \pi}\pvint_{4M^2}^\infty ds'\,{1\over s'-s} X(s')R(s')
\quad  X(s')=\tan\delta(s').
\en
It is useful to  split the integration region into several
sub-intervals in order to  accommodate fast variations of the
integrand (we have used up to seven intervals in our numerical work). 
Then, every sub-interval $[a,b]$ is mapped to $[-1,1]$ and the
the quantity $ X(s') R(s')$ is expanded over a basis of Legendre polynomials.
This allows us to perform the principal value integration in \rf{onedim} 
using the exact formula\cite{abramowitz},
\be
\int_{-1}^1 du\,{P_L(u)\over u-z}=2 Q_L(z)
\en
Here $Q_L(z)$ is the so-called Legendre function of the second kind. It is
crucial, in order to ensure the success of the calculation, that it be
computed to very high accuracy.  An algorithm, based on using the
recursion relations in the forward direction if $\vert z\vert<1$ and in the
backward direction otherwise, proves adequate. One obtains a discretised
approximation to the integral over $[a,b]$
\be
\pvint_a^b ds'{1\over s'-s_k} X(s') R(s')\approx
\sum_{i=1}^N \hat W_i\left[1+{2(s_k-b)\over b-a} \right] X(s_i) R(s_i)
\en
where
\be
s_i={1\over2}(a+b+(b-a)u_i),\quad
\hat W_i[z]=-w_i\,\sum_{j=0}^{N-1}
(2j+1)P_j(u_i)Q_j\left(z\right)\ ,
\en
and $u_1,\ldots ,u_N$ are the set of N Gauss-Legendre integration points
(i.e. the zeros) of $P_N(u)$ and $w_1,\ldots ,w_N$ are the associated set of
weights. In the case where $b=\infty$ (last sub-interval), we use
\be
s_i={2a\over 1-u_i}
\en
and 
\be
\pvint_a^\infty ds'{1\over s'-s_k} X(s') R(s')\approx
{2a\over s_k}\sum_{i=1}^N \hat W_i\left[1-{2A\over s_k} \right]
{X(s_i) R(s_i)\over 1-u_i}
\en
In this manner, the functional equation for the function $R(s)$ gets
transformed into a set of $M$ linear equations for $R(s_1),\ldots ,R(s_M)$
where $M=nN$, $n$ being the number of sub-intervals. We note that this is a
homogeneous system which, strictly speaking, has no nontrivial solution
unless the determinant vanishes. In practice, it does not exactly vanish.
It is only in
the limit of $N\to\infty$, in fact, that the determinant vanishes. 
In addition, one
wants to specify the value at zero $R(0)$ and this  generates one additional
equation, which is non-homogeneous. A solution can be defined by dropping
one of the homogeneous equations.
A numerically stable way of performing this, is to use
the singular-value decomposition of the $(M+1)\times M$  
matrix of the linear equation system\cite{recipes}.

We have performed several checks of the numerical calculations: a) we have
verified that upon using the T-matrix of the Truong-Willey type \cite{TW}
the analytical result was accurately reproduced  b)we have verified
the stability of the result when varying the number of integration points up
to several hundred points and c) we have also verified that the determinant of
the n-independent solutions obtained numerically accurately reproduces the
result which is known analytically (see sec. 4.C).

\newpage
\begin{figure}[ht]
\begin{center}
\title{\bf Figure 1}
\includegraphics*[scale=0.8]{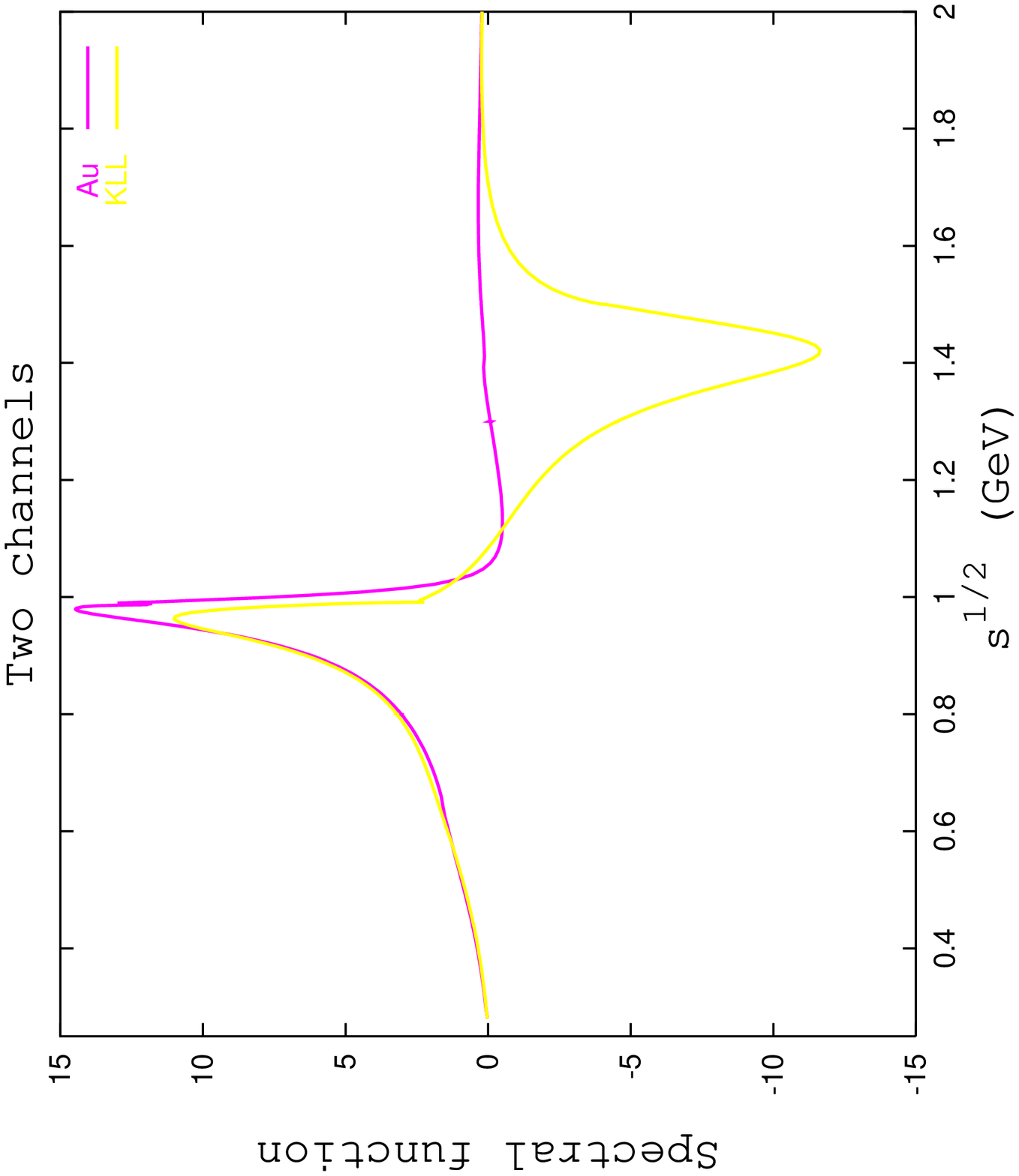}
\caption{\sl
Spectral function computed from T-matrix models with two coupled channels.
The solid line corresponds to the T-matrix of ref.\cite{au}, the dashed line
corresponds to ref.\cite{kll1}
}
\end{center}
\end{figure}

\newpage
\begin{figure}[ht]
\begin{center}
\title{\bf Figure 2}
\includegraphics*[scale=0.8]{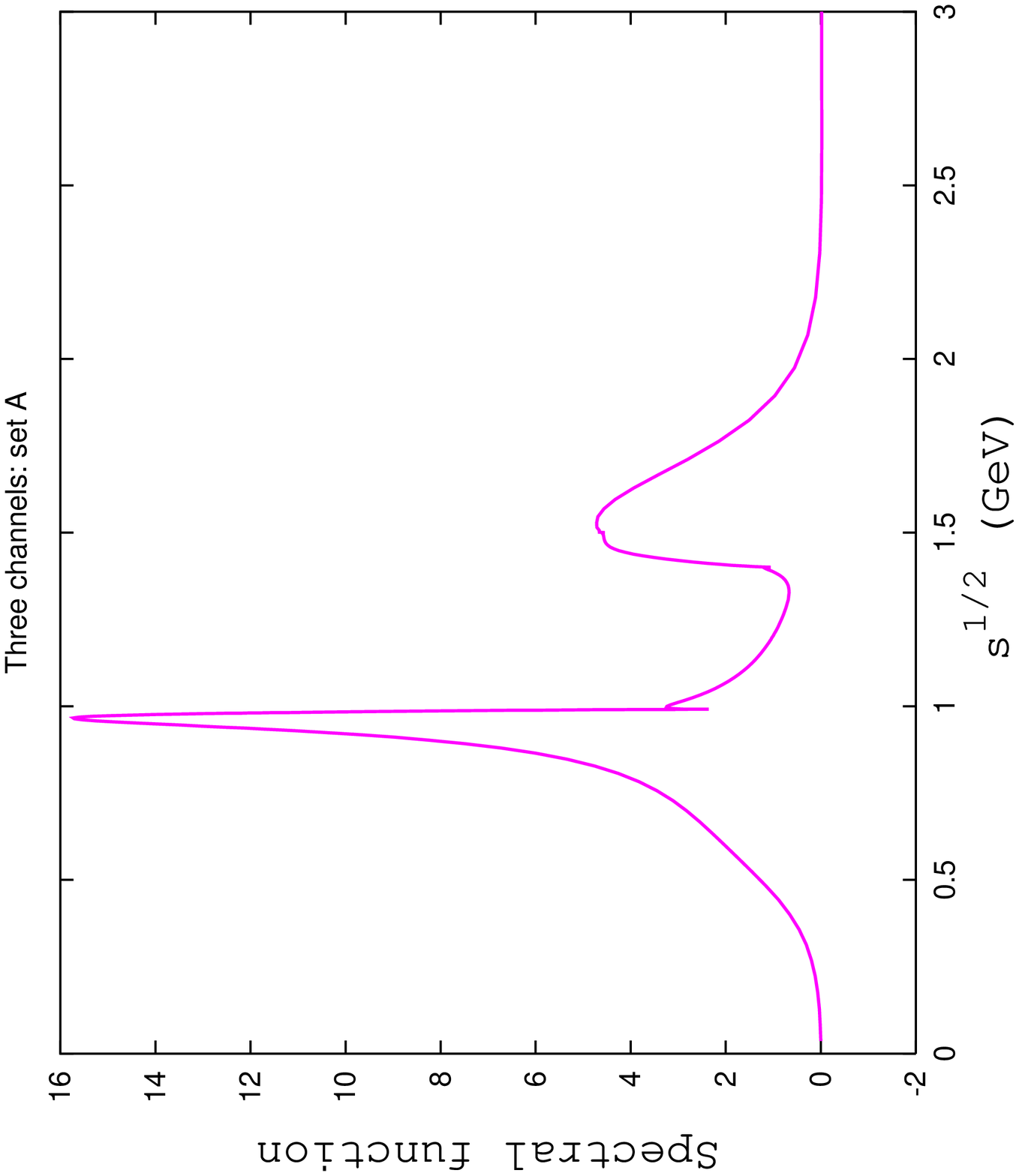}
\caption{\sl
Spectral function computed from the 3-channel T-matrix
model of ref.\cite{kll1}. The curve corresponds to the parameter set A of
this reference
}
\end{center}
\end{figure}

\newpage
\begin{figure}[ht]
\begin{center}
\title{\bf Figure 3}
\includegraphics*[scale=0.8]{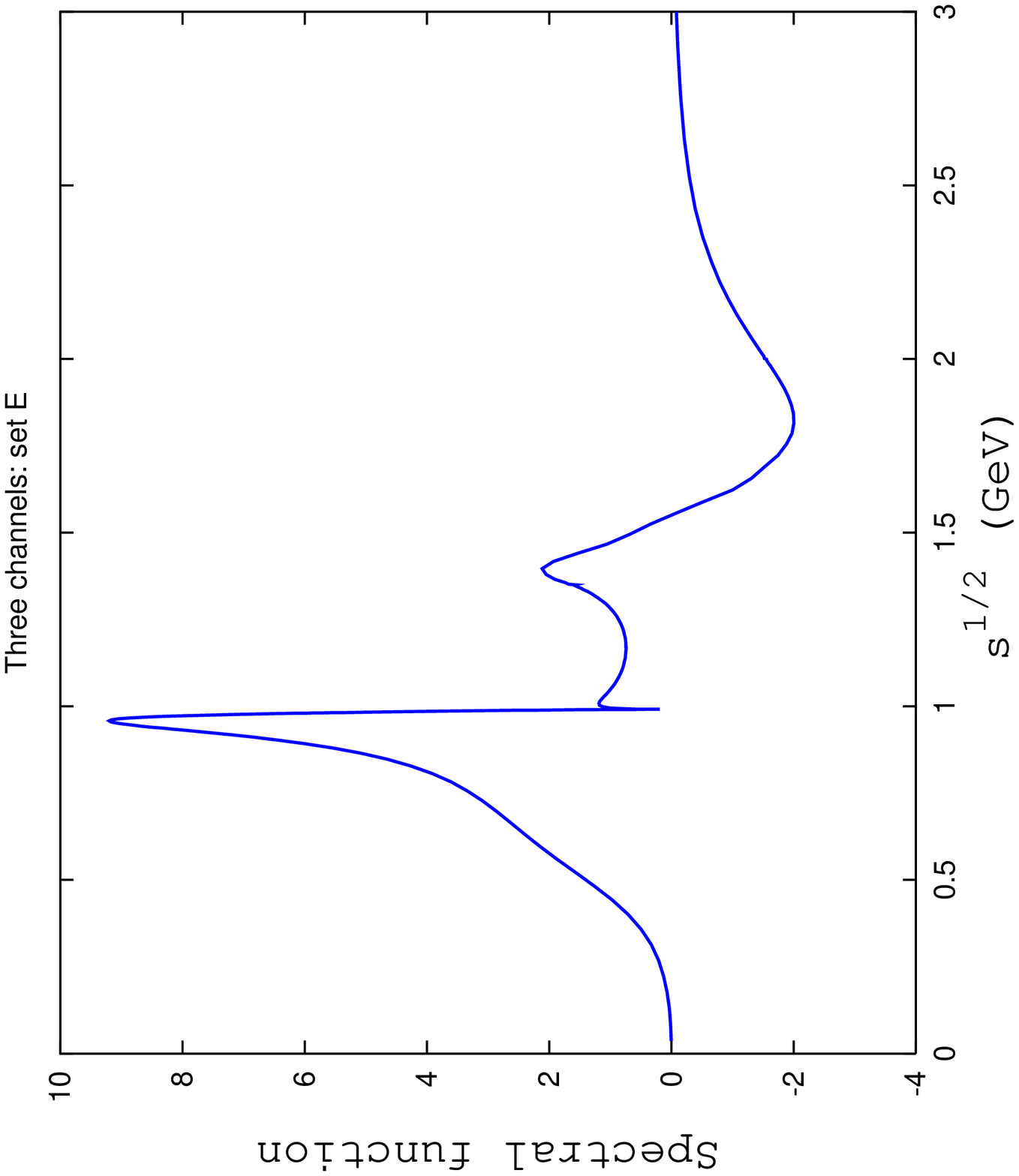}
\caption{\sl
Spectral function computed from the 3-channel T-matrix
model of ref.\cite{kll2}. The curve corresponds to the parameter set E of
this reference
}
\end{center}
\end{figure}

\newpage
\begin{figure}[ht]
\begin{center}
\title{\bf Figure 4}
\includegraphics*[scale=0.8]{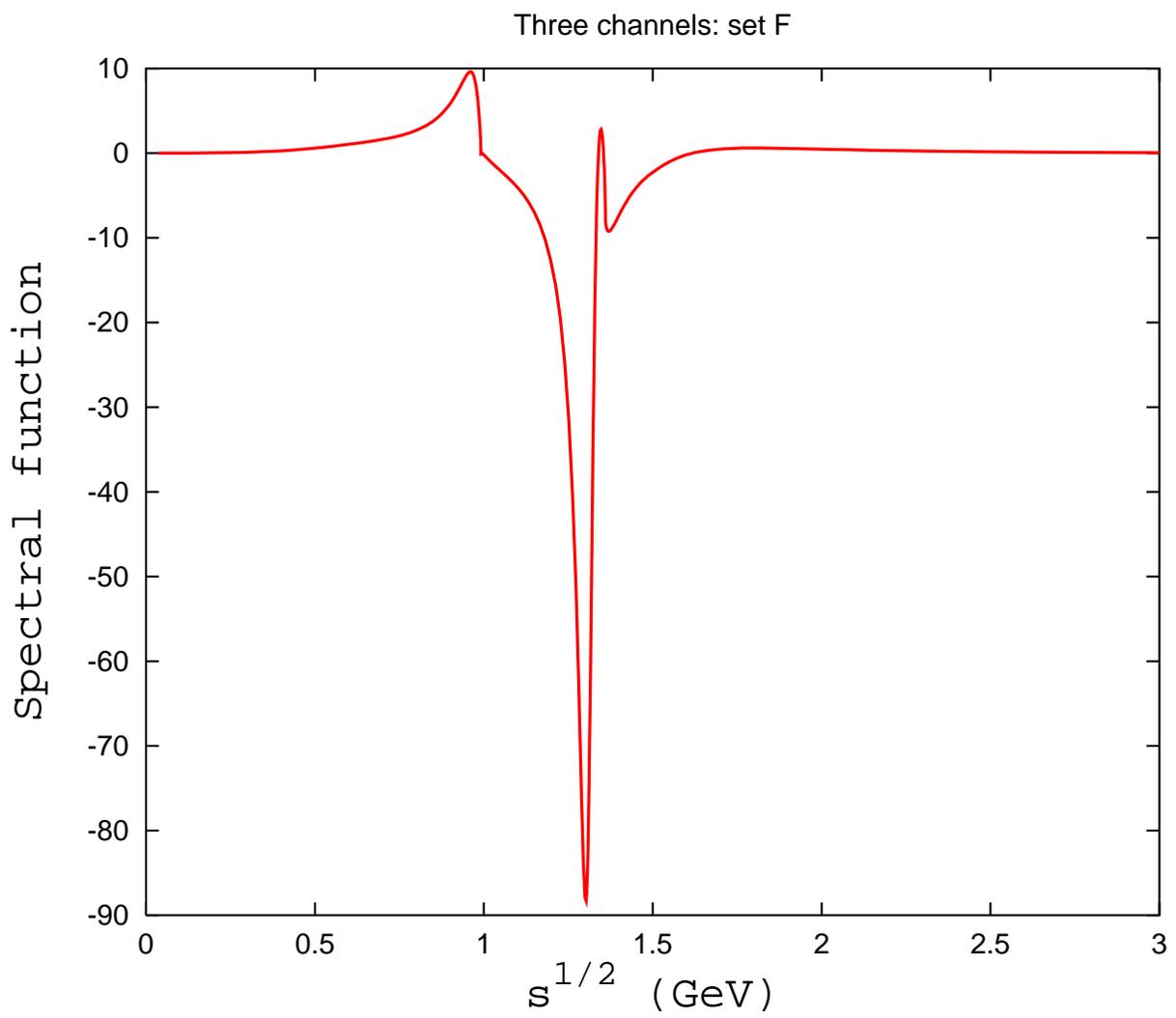}
\caption{\sl Same as Fig.3. The curve corresponds to the parameter set F.
}
\end{center}
\end{figure}
\end{document}